\documentclass[twocolumn,prl]{revtex4-1}
\usepackage[dvips]{graphicx}
\begin{document}

\title{Recursive Quantum Repeater Networks}

\author{Rodney Van Meter$^{1}$, Joe Touch$^2$, Clare Horsman$^3$}
\affiliation{$^1$Faculty of Environment and Information Studies, Keio University rdv@sfc.wide.ad.jp, http://web.sfc.keio.ac.jp/\char'176 rdv/\\
$^2$ University of Southern California/Information Sciences Institute touch@isi.edu\\
$^3$ Shonan Fujisawa Campus, Keio University clare@sfc.wide.ad.jp}
\date{\today}

\begin{abstract}%
  Internet-scale quantum repeater networks will be heterogeneous in
  physical technology, repeater functionality, and management.  The
  classical control necessary to use the network will therefore face
  similar issues as Internet data transmission.  Many scalability and
  management problems that arose during the development of the
  Internet might have been solved in a more uniform fashion, improving
  flexibility and reducing redundant engineering effort.  Quantum
  repeater network development is currently at the stage where we risk
  similar duplication when separate systems are combined.  We propose
  a unifying framework that can be used with all existing repeater
  designs. We introduce the notion of a Quantum Recursive Network
  Architecture, developed from the emerging classical concept of
  \emph{recursive networks}, extending recursive mechanisms from a
  focus on data forwarding to a more general distributed computing
  request framework.  Recursion abstracts independent transit networks
  as single relay nodes, unifies software layering, and virtualizes
  the addresses of resources to improve information hiding and
  resource management.  Our architecture is useful for building
  arbitrary distributed states, including fundamental distributed
  states such as Bell pairs and GHZ, W, and cluster states.
\end{abstract}

%
\maketitle


  



%
%

\newcommand{\ket}[1]{\ensuremath{|#1 \rangle}}
\newcommand{\bra}[1]{\ensuremath{\langle #1|}}
\newcommand{\braket}[2]{\ensuremath{\langle #1|#2 \rangle}}
\newcommand{\ketbra}[2]{\ensuremath{|#1 \rangle \langle #2|}}
\newcommand{\ro}[1]{\ensuremath{|#1 \rangle \langle #1|}}
\newcommand{\av}[1]{\ensuremath{\langle #1 \rangle}}
\newcommand{\real}{\ensuremath{\mathrm{Re}}}
\newcommand{\trace}{\ensuremath{\textsf{Tr}}}


\section{Introduction}

\emph{Communication} is the sharing of data -- bits, in modern
digitial systems -- between a pair of endpoints, with a certain
probability of the shared data being the same at both ends.
Shannon~\cite{shannon} addressed this problem for the case of a simple
\emph{channel}, where the endpoints are known \emph{a priori}.  In the
sixty years since his seminal analysis, researchers have extended the
concepts to apply to networks consisting of more than a single sender
and receiver~\cite{tanenbaum:networks-4th-ed}.  Except for the
physical means of creating entanglement, quantum repeater networks
require solutions to the same problems as all classical communications
networks, allowing us to reuse many of the engineering principles
developed for classical networks.

Quantum repeater networks support the sharing of quantum states.
Distributed, entangled states of various kinds are used in distributed
decision
algorithms~\cite{tani05:_quant_leader_elect,ben-or2005fast,gaertner:qbyz-demo},
the creation of shared, secret random
numbers~\cite{ekert1991qcb,markham2008graph}, distributed
arithmetic~\cite{van-meter07:_distr_arith_jetc}, secure, distributed
function computation~\cite{crepeau:_secur_multi_party_qc}, quantum secret
sharing~\cite{hein2006egs}, physical operations such as remote
synchronization of clocks~\cite{chuang2000qclk}, and a range of other
applications~\cite{buhrman03:_dist_qc,dhondt05:_dist-qc,lalire2004paa,van-meter08:_quant_internet_apps}.

In this paper, we examine the problem of delivering distributed,
entangled states across a large, topologically complex,
technologically diverse internetwork, owned and operated by many
different organizations and deployed over a period of years.  These
problems will be particularly evident as a communication request
crosses the boundary between two networks based on different
entanglement purification and forwarding schemes.  We consider what
lessons may be learned from the historical development of the
Internet. In order to prevent similar problems from occurring in a
quantum repeater Internet, we propose a solution based on classical
recursive network architectures, extended to support requests for
distributed computation.  As our goal is planetary-scale networks
composed of perhaps billions of nodes, scalability is a critical
feature and motivates our work.

Researchers have proposed a ``quantum Internet'', using the term
``Internet'' as a synonym for a shared, global
network~\cite{kimble08:_quant_internet,lloyd2004iqi}.
The term ``internetwork'' encompasses a richer meaning, literally
connecting multiple disparate networks.  In any large network that
grows and evolves over time, multiple different technologies and many
different management domains will coexist, forming an internetwork or
\emph{internet}.  This organic growth and the sheer scale of large
networks pose several categories of problems: (1) ensuring
interoperability among technologies that are heterogeneous (at both
the physical and logical levels); (2) reconciling the competing needs
and policies of independent organizations (including the desire to
keep information about the network internals private); (3) choosing a
technical approach for the routing, naming, and resource discovery
problems that is robust in the face of this heterogeneity and
federated operation; and (4) managing communication requests using
incomplete, out-of-date information about the dynamic state of the
network, including availability of resources and topological changes
occurring as nodes join and leave, and network links go up and down.

The evolution of the Internet provides a guide to designing
large-scale systems.  It is a hierarchical system, with the
communication functionality divided into a set of {\em protocol
  layers}, but not a truly recursive architecture (see
Sec.~\ref{sec:c-net}).  As the Internet grew from its predecessors,
various routing protocols were devised, and a two-layer hierarchical
structure created (using what are known as {\em interior gateway
  protocols} (IGPs) and {\em exterior gateway protocols} (EGPs)) to
hide internal topologies and provide scalability and management
autonomy and privacy.  This hierarchy has other layers, but until
recently the number and structure of layers has been fixed.  Over
time, it became desirable to build one network on top of another, or
to translate addresses as data packets cross network boundaries.
Virtual networks, tunnels, overlays, mobile IP, and network address
translation (NAT) are used to achieve various technical and
operational goals, but all interfere with the original, uniform scheme
for addressing and routing, and have resulted in much duplicated
engineering effort as functionality is re-implemented at various
levels~\cite{pna,tanenbaum:networks-4th-ed}.  Additional layers of
this hierarchy have been added, using encapsulation-based subnets,
such as LISP~\cite{meyer2008locator}. Recently, these layers have been
understood as instances of a more general and flexible recursive
architecture that supports arbitrary layering. Our goal is to apply
that more general approach to quantum networking now, without
restricting technological innovation or organizational choice, while
supporting the quantum networking protocols already under development.

An emerging concept in classical networking is \emph{recursive
  networks}~\cite{pna,rna-tr}, in which a subset of a network can be
represented as a single node at a different layer of that network.
Recursive networking is used to unify multihop forwarding, embedded
topologies and other forms of virtualization, and the software
layering common in network protocol stacks.

\begin{figure*}[t]
  \begin{center}
    \includegraphics*[width=120mm]{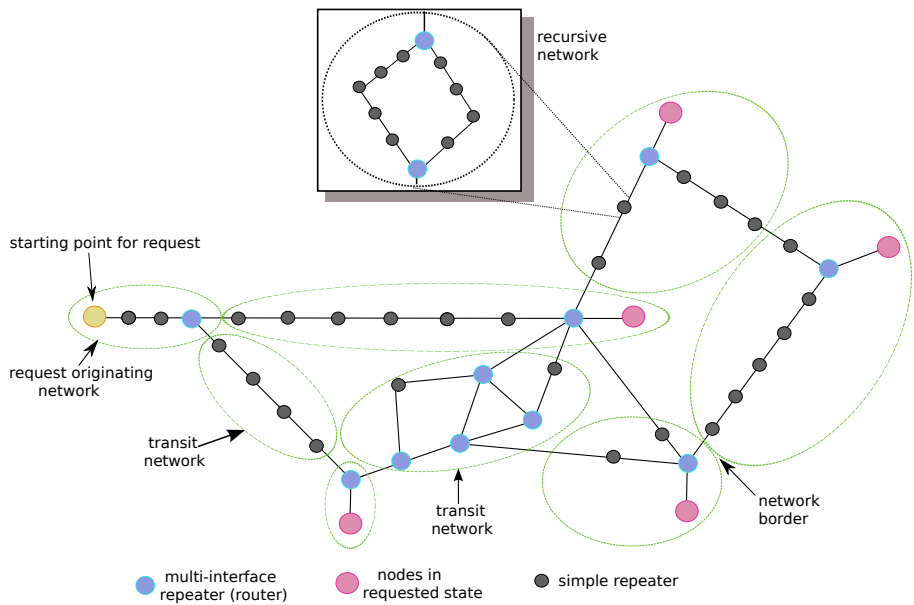}
  \end{center}
  \caption{Large-scale quantum repeater networks will consist of many
    nodes.  Many long-line connections between routing-capable
    vertices (``routers'') will consist of multiple-hop chains of
    simpler repeaters.  In a recursive network, each node in the
    figure can actually represent a complete network itself that
    provides the services of a single node.}
  \label{fig:network}
\end{figure*}

Recursive networks allow individual networks to offer data forwarding
services without requiring requesters to understand the detailed
topology or technology of the network, and for that process to be
repeated at multiple layers in the network more cleanly than the
two-level EGP/IGP system.  The networks that may form part of the
path, but do not include nodes that are part of the requested state,
are \emph{transit networks}.  To external requesters, a transit
network appears as a single node.  Internally, nodes within the
network can in turn be networks, in recursive fashion.
Fig.~\ref{fig:network} shows an example network topology.  The
Internet and telephone network both exhibit a fixed set of layers, but
recursive approaches are more general and scalable.  Such approaches
have only recently been applied in the Internet, in
LISP~\cite{meyer2008locator} and Rbridges~\cite{perlman2004rbridges},
but adoption of the concept earlier in their evolution may have
alleviated some of the above scalability problems.  Recursive networks
will be described in more detail in Sec.~\ref{sec:c-net}.

Our engineering philosophy is to adapt classical solutions to the
quantum domain where possible, an approach that has proven fruitful in
other areas of quantum systems engineering.  An excellent example is
quantum error correction (QEC), where many of the ideas originate in
classical error correcting codes (see Ref.~\cite{devitt09:_idiot_qec}
and Ref.~\cite{nielsen-chuang:qci} and references therein).  Quantum
arithmetic likewise has built on classical concepts (see
e.g. Refs.~\cite{vedral:quant-arith,beckman96:eff-net-quant-fact,feynman:lect-computation,van-meter04:fast-modexp,draper04:quant-carry-lookahead}).
Both fields are rich areas of study, resulting in both new ideas and
rigorous, creative applications of old ones; we expect the same to be
true as we engineer quantum networks.

We have chosen to adopt recursive networking as a framework for
quantum networks.  The principal difference between classical and
quantum networking lies in the semantics of the request.  In a
classical network, the arrival of a data packet represents an implicit
request to forward that packet on toward its specified destination.
Quantum networks (e.g., those based on teleportation) transmit not
quantum data itself, but requests for the execution of operations
which will {\em recreate} quantum states extinguished at other sites,
or create new distributed, entangled states.  Thus, {\it quantum
  requests are more complex than classical ones, and can no longer
  remain implicit}.  Recursive quantum networks require significant
extension of the classical concepts to offer distributed state
creation services based only on a description of the desired state (or
the computational steps necessary to create the desired state).

Our quantum recursive network architecture (QRNA) contributes to
solving all four of the scaling problems, and will circumvent the
evolutionary problems above.
Ref.~\cite{van-meter07:banded-repeater-ton} defines a protocol stack
for purify-and-swap repeaters (see Sec.~\ref{sec:q-net}) with separate,
but stackable, protocol layers for entanglement swapping and one
specific purification mechanism, building long-distance Bell pairs.
QRNA generalizes and extends this to a more complete architecture for
creating arbitrary distributed quantum states spanning heterogeneous,
autonomous networks.  In this paper, we describe the recursive
framework, identify the contents of the necessary messages, show how
they can be executed recursively, and enumerate the operational and
architectural benefits.

After opening with networking background, both quantum
(Sec.~\ref{sec:q-net}) and classical (Sec.~\ref{sec:c-net}), we
describe the request structures that make the recursive architecture
possible (Sec.~\ref{sec:rec-q-reqs}), then show how this concept makes
real-world deployment of truly large-scale, heterogeneous networks
practical (Sec.~\ref{sec:impl}).

\section{Quantum Networking}
\label{sec:q-net}

To date, most research on quantum networking has focused on systems
for creating high-fidelity generic entangled states, such as
end-to-end Bell pairs~\cite{dur:PhysRevA.59.169} or larger graph
states on model networks~\cite{benjamin:njp-7-1-194,munro05:_weak}.
The resulting generic states are then used for the remote execution of
quantum gates~\cite{gottsman99:universal_teleport}, teleportation of
valuable application-level
qubits~\cite{bennett:teleportation,furusawa98,S.Olmschenk01232009}, or
the creation of shared classical random bits via measurement (e.g.,
for quantum key distribution
(QKD)~\cite{ekert1991qcb,markham2008graph}).  The more general concept
is the creation of arbitrary distributed, entangled states; thus, a
quantum network is effectively a large-scale distributed quantum
computing system.

The problems that must be solved to implement quantum networks include
loss and fidelity degradation in optical channels, as well as network
issues such as routing and resource
management~\cite{satoh10:qdijkstra-aqis,aparicio10:muxing-aqis}.  The
issue of an imperfect but abstract channel was the first attacked in
the literature, so we will begin our discussion there.

Over a channel between sender and receiver, the goal has been to
distribute entangled states without significant fidelity degradation.
Even over short distances the fidelity will degrade, and some signals
may also become lost; the fidelity and probability of successful
entanglement vary depending on the physical mechanism and whether
single photons or stronger signals are
used~\cite{childress05:_ft-quant-repeater,ladd06:_hybrid_cqed,munro:PhysRevLett.101.040502,van-loock06:_hybrid_quant_repeater}.

The standard solution has been to establish multiple lower-fidelity
Bell pairs between adjacent nodes, and then consume some of them in a
{\em purification protocol} to leave the remaining links as close as
possible to the desired Bell
state~\cite{briegel98:_quant_repeater,dur2007epa,bennett1996pne,PhysRevLett.77.2818}.
Purification protocols can be run as many times as is required,
consuming lower-fidelity Bell pairs at each step.  Finally, a single
high-fidelity Bell pair is produced end-to-end in the network, with
the caveat that achievable fidelity remains capped by local gate,
measurement and memory errors.

Standard purification requires two-way classical communication between
the two ends of the Bell pair being purified, which can result in
delays that harm both fidelity and performance, perhaps
irreparably. There is also always a nontrivial possibility of failure
of the purification: all the entangled links are consumed, but a
high-fidelity Bell pair is not produced.  This information must be
shared between the end points.  Purification can also be done using
one-way communication and quantum error correction, with some loss in
channel capacity~\cite{dur2007epa}, and can also be performed on graph
states larger than two qubits~\cite{dur2003mep,kruszynska2006epp}.

The solution to the problem of signal loss (exponential in distance
for individual photons in optical fiber) has been to use
\emph{repeater nodes}, proposed to be placed at short intervals
(generally of the order of tens of
kilometers~\cite{childress05:_ft-quant-repeater,childress2006ftq,van-loock06:_hybrid_quant_repeater}).
With either a chain of repeaters or a network comes the problem of
moving data over multiple hops.

The obvious solution for moving quantum data is to simply forward a
qubit from node to node using
teleportation~\cite{bennett:teleportation}.  However, this hop-by-hop
approach has long been considered unworkable in realistic environments
because imperfect local gates and memories result in unacceptable
degradation of fidelity~\cite{dur:PhysRevA.59.169,hartmann06}.  This
limitation led to the proposed use of purification over multiple
hops~\cite{dur:PhysRevA.59.169,dur2007epa}.  To create the end-to-end
entangled state, teleportation is often proposed to be used to perform
{\em entanglement swapping}, often in a nested manner that doubles the
span of entanglement at each step.  With recent advances in the
distributed use of error
correction~\cite{PhysRevA.79.032325,fowler2008htu} and operational
tactics for repeaters~\cite{munro2010quantum}, additional options are
available.  Perhaps the principal architectural choice facing the
network architect is whether to attempt to revive the hop-by-hop
approach, to operate in a distributed fashion (the purify-and-swap
approach), or to take a radically different approach to moving data
through the network, such as using the surface
code~\cite{fowler2008htu,raussendorf07:_2D_topo,raussendorf07:_topol_fault_toler_in_clust,wang2009ter}
or other quantum error correcting codes.

Recent development of such new approaches started with
Jiang \emph{et
  al.}~\cite{PhysRevA.79.032325}, where error correction is used to replace some
amount of purification, particularly in the context of long-distance
QKD.  Calderbank-Shor-Steane
(CSS)~\cite{calderbank96:_good-qec-exists,steane96:_qec} encoding is
used to replace long-range purification, although Bell pairs over
single hops still require purification and the associated classical
messaging, both between adjacent nodes and across the network in the
case of failure events. More recently, Munro \emph{et al.} have
presented a scheme where purification is replaced entirely with
quantum error correction and parallelization of low-fidelity link
generation~\cite{munro2010quantum}. Instead of consuming multiple
low-fidelity Bell pairs in a purification protocol, they show how a
single high-fidelity Bell pair can be encoded in these multiple pairs,
and give the explicit example of a repetition code. Only a single
link-level classical message needs to propagate backwards from the
receiver to the sender and, most importantly, the encoding is
deterministic. There are therefore no end-to-end entanglement or
purification failure messages that require propagation through the
network.

In addition to using error correction codes to protect Bell pairs as
they propagate, other recent work demonstrates that data qubits
themselves can be transported in this way. Fowler \emph{et al.}
propose a scheme using the topological surface code on a regular
cluster state superimposed on a repeater
network~\cite{PhysRevLett.104.180503}. Each node contains a block of
surface code, with the block edges coupled through the entangled links
between nodes. The data propagates through the network directly, rather
than being held until a high-fidelity Bell pair is available and then
teleported.

There are therefore many options available for managing and using a
repeater network. Either standard purification or error correction can
set up end-to-end Bell pairs. Error correction-based schemes can
transmit application data in a hop-by-hop fashion, without end-to-end
entanglement having ever been present at a given point in time. The hop-by-hop approach is attractive because of its similarity to
classical networks, although a full accounting of resources used,
including error correction to protect the transmitted states while
they are operated on at the transit nodes, remains to be done.  In the
purify-and-swap case, activity must be coordinated with numerous
waypoints or rendezvous points via classical messages, which, prior to
the QRNA architecture proposed here, required explicit, static
assignment of those points.  In other architectures, the requirements
vary from complete control of activity on an entire path to a
send-it-and-forget-it approach similar to that of packet forwarding on
the classical Internet.  One of our goals is to create a request model that
will support interoperation of all these approaches.

\begin{figure}[t]
  \begin{center}
\includegraphics[width=50mm]{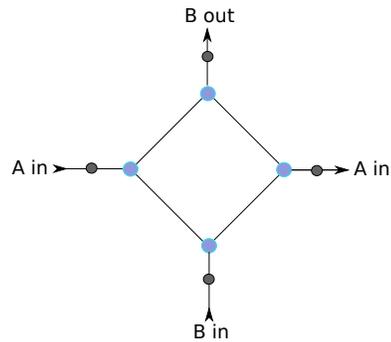}
  \end{center}
  \caption{Graph states are useful both at the application level and
    as communications channels; a diamond junction like this one can
    be used to send two qubits simultaneously from place to place in
    measurement-based quantum computing.}
  \label{fig:diamond-graph}
\end{figure}

While the main focus of research into quantum repeaters has been for
2-party message transmission, it is also possible to use such a
network for distributed measurement-based quantum communication
protocols. These protocols use highly-entangled graph states to
propagate information through the use of intermediate measurement,
consuming the graph state resource in the process.  Such schemes can
propagate either application data or half of a Bell state to generate
end-to-end entanglement. Managed properly, multiple data movement
requests can also be satisfied from the same graph state, as in
Fig.~\ref{fig:diamond-graph}~\cite{raussendorf03:_cluster-state-qc}.
As given, such scheme do not deal with realistic (impure) states, and
will require adaption to an error correction/purification-based
network. However, the key idea is already present in recent schemes,
that a quantum network where data propagates directly is capable not
only of hop-by-hop communication, but also distributed computation
without end-to-end classical control messaging~\cite{ballistic}.

It is obvious from this discussion that repeater nodes in a quantum
network differ significantly from classical signal repeaters. They are
not classical signal amplifiers; they provide base-level entanglement
with neighbors, data transmission services and the quantum computation
and classical communication operations that are necessary for both
purification and entanglement swapping.  They are small,
limited-functionality quantum computers in their own right, and
fulfill the role of routers in the Internet.  We will take advantage
of these computational capabilities in QRNA, as will be described in
Sec.~\ref{sec:rec-q-reqs}.  First, we preview some of the key ideas of
QRNA while reviewing some of the principles of networking.

\section{Computer Networks}
\label{sec:c-net}

This section explains how quantum networking builds on concepts from
classical networking, notably the emerging concept of \emph{recursive
  networks}.

\subsection{Network Topology}

A \emph{network} is a group of interacting parties, where each of the
members potentially wants to interact with any of the other members.  A
computer network consists of \emph{nodes}, which are typically a form
of computer, and \emph{links}, which carry messages from node to node.
A node uses a \emph{physical interface} to connect to the link.
Links may be bidirectional or unidirectional, and bidirectional links
can be further divided into full- and half-duplex, depending on
whether they support concurrent bidirectional transfer or must be time
multiplexed.

In classical networks, links are also described as 2-party or
multiparty.  Two-party links involve two known, fixed nodes;
multiparty links involve multiple nodes, in which the set of nodes may
change over time (i.e., membership is not known, but rather
discovered), and where a single message can be received by an
individual node, or by all or some of the nodes (known as broadcast or
multicast, respectively).  The single-receiver case is equivalent to a
bus, and numerous physical implementations of quantum systems support
addressing individual qubits in such a fashion, generally through a
shared waveguide or resonator (see, for example, Refs.
\cite{kielpinski:large-scale,ladd10:_quantum_computers,spiller05:_qubus,steane97:ion-trap,van-meter10:dist_arch_ijqi}).

Classical broadcast or multicast copies data to all interested
receivers.  The direct equivalent for quantum communication would use
GHZ states~\cite{greenberger89:ghz} for the FANOUT creation of
GHZ-like states, giving each node a part of the state.  In FANOUT \cite{fanout}, a
single qubit $\ket{\psi} = \alpha \ket{0} + \beta \ket{1}$ is expanded
into an $N$-qubit GHZ-like state
\begin{equation}
\ket{\psi'} = \alpha\ket{0}^{\otimes N} + \beta\ket{1}^{\otimes N},
\label{eq:ghz}
\end{equation}
where $\alpha$ and $\beta$ conform to the usual normalization
condition $|\alpha|^2 + |\beta|^2 = 1$ and the notation
$\ket{0}^{\otimes N}$ indicates the $N$-qubit state $\ket{0}\otimes\ket{0}\ldots\otimes\ket{0}$.  In distributed algorithms, this gives each node
in the network access to the same quantum variable, which can be used
in further quantum computations.  This FANOUT of arbitrary quantum
information allows for quantum computation to proceed in parallel
(e.g., when used to distribute carry information in
arithmetic~\cite{van-meter07:_distr_arith_jetc}).  Generic GHZ states
are also used to make coordinated, distributed
decisions~\cite{tani05:_quant_leader_elect}.

Other multi-party entangled states, such as W and graph states, will
be used in different fashions in quantum algorithms.  These entangled
states may be created using some direct, multi-party physical
interaction~\cite{yamaguchi05:_weak-nonlin-qec,choi10:4-qubit-photon-w-state},
or more likely by local creation of the complex states that are
propagated outwards using two-party distributed Bell pairs.  This
tradeoff is discussed further below.

As a result, quantum networks correspond most closely to classical
networks with 2-party unidirectional links, with the primary remaining
difference being the data being communicated (qubits \emph{or} larger
entangled states, and the long-distance gates necessary to execute
them) and the means by which data is relayed.

\subsection{Multihop Communication}

\begin{figure}[t]
  \begin{center}
    \includegraphics[width=8cm]{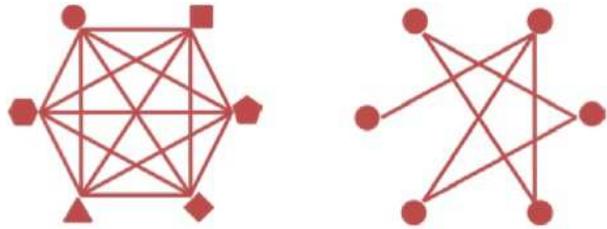}
  \end{center}

  \caption{Multiparty communication without forwarding requires $N^2$ links (left), but can use only $N$ links (right) with forwarding.}
  \label{fig:forwarding}
\end{figure}

As noted above, quantum links are two-party, and the endpoints are
known. This is the basis of Shannon's model of communication, but it
requires $N^2$ links (given $N$ nodes in a network) (see
Fig.~\ref{fig:forwarding}, left side). More commonly, such full
connectivity is supported with a smaller set of links using multihop
forwarding (see Fig.~\ref{fig:forwarding}, right side).  Forwarding is
the fundamental concept that enables scalability in physical systems,
both in distance and number of nodes: individual nodes have multiple
physical interfaces, receive messages, and make decisions about how
best to send the messages on to their respective destinations.

Given two nodes that want to communicate, a path must be found along
the existing links, and communication along that path cascaded. In
classical networks this is called \emph{packet forwarding}, and is the
basis of the Internet. There are other steps involved in classical
networking, such as name resolution (finding the location of a named
item), and routing (finding a network path to a location); we assume
they function here as they would in any classical network, and address
these issues further in Sec.~\ref{sec:state-name}.

\begin{figure}[t]
  \begin{center}
    \includegraphics[width=8cm]{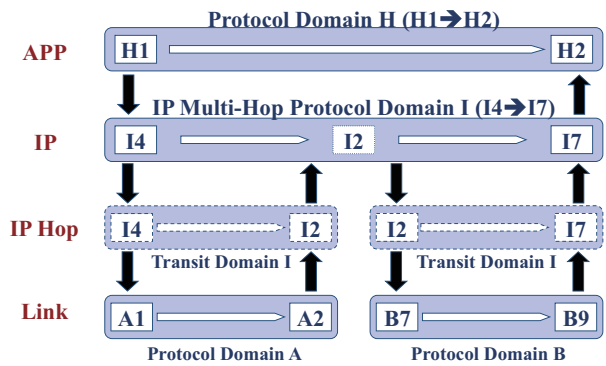}
  \end{center}

  \caption{Classical recursive multihop/multilayer architecture,
    including forwarding steps (Hop) as part of a multihop path (IP,
    here). APP is the application layer, IP is the Internet Protocol
    layer that connects different link networks, and the link layer
    provides a physical connection over a single hop (e.g., dial-up),
    or by a similar multihop network (not shown, e.g., Ethernet).}
  \label{fig:layering}
\end{figure}

In modern communication architectures, the functionality is divided
into a set of \emph{protocol layers}, as shown in
Fig~\ref{fig:layering}.  Each layer has a different role in supporting
the end-to-end communication requested by the originating
application. Each layer uses services provided by its lower layers to
communicate with its peer layer at the remote node. In practice, the
software implementing these layers may be integrated into a single
module, but they are commonly described as if they were implemented
separately.

Quantum networking is described well by the emerging concept of
\emph{recursive networks}. Recursive networking was developed in 2000
to describe multi-layer virtual networks that embed networks as nodes
inside other networks~\cite{dynabone}, and has since evolved as a
possible architecture for the future
Internet~\cite{pna,rna-tr,net-ipc,dynabone,druid,rna}. It has been
used to unify the layering of protocol software, message forwarding,
and topology embedding. Classical message forwarding is explained by a
kind of \emph{tail recursion}~\cite{steele:tail}, in which the last
step performed in the operation is the recursion itself. Tail
recursion is basically iteration accomplished using recursion, where
instead of each step recursing (and thus pushing information on a
stack) and the final step popping the entire stack, each step
overwrites the top of the stack so that the last step can complete
more efficiently.  Quantum message forwarding is closer to topology
embedding, as in the ``recursive router'' concept introduced in the
X-Bone~\cite{dynabone}.

\subsection{Recursive Networking}

\begin{figure}[t]
  \begin{center}
    \includegraphics[width=8cm]{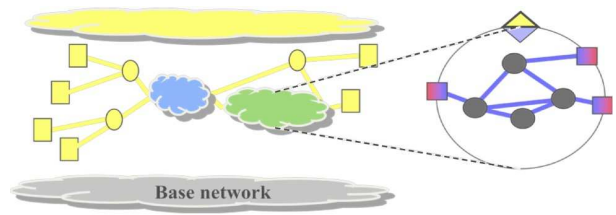}
  \end{center}

  \caption{Classical network recursion, showing a subnet (cloud) that
    acts as a router in the overlay (yellow), where the recursive
    router (inset circle) consists of ingress and egress virtual hosts
    (shaded squares on the inset) and interior routers (gray
    circles).}
  \label{fig:recursion}
\end{figure}

A subset of a network can be embedded in the overall topology; such
embeddings are useful in classical networks to hide complex subnet
structure from being visible to the overall network (see
Fig.~\ref{fig:recursion}). This subnetwork embedding is called
\emph{recursive networking}, and can be part of a broader approach to
network architecture~\cite{net-ipc,druid,rna}.

In classical networking, nodes acting as message sources or sinks are
called \emph{hosts}, and nodes acting as relays are called
\emph{routers}. A recursive network represents the embedded subnet as
a router at the higher level, where the ingress and egress nodes of
that subnet act as hosts inside the subnet (see
Fig.~\ref{fig:recursion}, detail). Such embedding can happen many
times, sometimes on top of existing embeddings, which is why this is
called \emph{recursive}.  Note the similarity of
Fig.~\ref{fig:recursion} to Fig.~\ref{fig:network}.  When a given
layer in the protocol stack provides an interface to its clients that
is identical to the interface of the services it uses (e.g., packet
forwarding to nodes in a particular namespace), implementation of
recursion is straightforward.

Classical networking uses recursion to represent topology hiding, but
we can also consider the entire network architecture as recursive as
well~\cite{pna,rna-tr}. As an architectural principle, recursive
networking explains layering of protocols (and their modular software
architecture), name resolution, routing, and forwarding as more than
just artifacts of the current Internet~\cite{druid}.

In most forms of quantum networking, recursion as a request moves
through the network is true recursion; we cannot transform it into a
hop-by-hop iteration and optimize it as tail recursion.  Fig. 2 of
Ref. \cite{van-meter07:banded-repeater-ton} shows how layers of quantum
repeaters are used to compose a sequence of individual hops into a
single, longer hop. This is the same composition process of tail
recursion in classical recursive networks, except that the layering is
left in place rather than collapsed as a simplification.

\begin{figure}[t]
\begin{verbatim}
deliver(data, src, dst) {
  process(data) -> newdata
  WHILE (here != dst) {
    found = FALSE
    FOREACH (lowerlayer) {
      map(src,dst,lowerlayer) -> newsrc, newdst
      IF (deliver(newdata, newsrc, newdst)
          == TRUE) {
        found = TRUE
      }
    }
    IF (found == FALSE) {
      /* if you get here, you failed to deliver 
         the data */
      FAIL 
    }
  }
  /* if you get here, you're at the
     destination */
  RETURN TRUE
}
\end{verbatim}
  \caption{The algorithm for recursive resolution and forwarding,
    adapted from Ref.~\cite{rna-tr}.  This algorithm is executed at
    each node as it receives data to be delivered.  {\tt src} and
    {\tt dst} are the source and destination addresses, and {\tt
      lowerlayer} refers to e.g. the layer on the receiving end of a
    downward arrow in Fig~\protect\ref{fig:layering}.}
  \label{fig:algorithm}
\end{figure}

Consider the steps of classical recursive networking, shown in
Fig.~\ref{fig:algorithm}.  When a packet is received by a node, the
packet is implicitly requesting that the node forward the packet on
toward its destination.  This algorithm is executed by the node to
perform the forwarding.  The {\tt process()} step may alter the
packet, including turning a single packet into more than one.  The
{\tt FOREACH} loop passes the packet(s) down the protocol stack, as in
Fig.~\ref{fig:layering}.

QRNA adopts a similar outline, with different semantics.  In quantum
networks, {\tt data} contains a computation request using virtual
identifiers for resources, and the {\tt process()} step represents the
local operations that are performed in a repeater toward fulfillment
of that request, such as the entanglement swapping that happens when
Bell pairs are spliced to form a longer
pair~\cite{dur:PhysRevA.59.169,van-meter07:banded-repeater-ton}.  The
output of {\tt process()}, {\tt newdata}, may be more than one
request.  The {\tt map()} function may modify the addresses in a given
request.  In our architecture, the node identifiers don't change
within a forwarding path, but requests may be retargeted from a
network destination to a node destination, as we will describe in
Sec.~\ref{sec:impl}.  The corresponding concept is shown in layered
communication, as supported by recursive networking, in
Fig.~\ref{fig:layering}.

Beyond this basic structure for forwarding, in QRNA the requests
themselves become recursive, and must be carried explicitly through
the classical network.  Next, we turn to the structure of these
requests.

\section{Recursive Quantum Requests}
\label{sec:rec-q-reqs}

Recursion is a natural model for quantum repeater networks because
purification, entanglement swapping, and Calderbank-Shor-Steane
(CSS)~\cite{calderbank96:_good-qec-exists,steane96:_qec} or surface
code error
correction~\cite{fowler2008htu,raussendorf07:_2D_topo,raussendorf07:_topol_fault_toler_in_clust,wang2009ter}
build on mixed, entangled states and produce other mixed, entangled
states, working toward a common goal of a high-fidelity, wide
area-distributed quantum state.  The similarity of the interfaces on
the top and bottom of a given protocol layer simplifies recursion,
allowing more or less arbitrary composition of protocol stacks.

In a large network (millions to billions of nodes spread across many
countries and organizations), direct management of the network as a
single, synchronous, shared, centrally managed system is impractical,
and even optimization of smaller portions of the network becomes a
computationally intractable combinatoric problem.  Applying recursion
abstracts away much of this complexity and allows us to effectively
manage the larger set of resources.  Each protocol layer, node or
network needs to recognize and be able to reach only a small subset of
the entire network's resources, and hides much of the underlying
complexity to allow its own clients to operate in a smaller subspace.

In order for recursion to be effective, we must have a well-defined
request-response model that allows us to combine protocol layers.
Before the requests and responses can be defined, we must be able to
name the distributed entangled states themselves, and the resources
comprising them.  The next two subsections address these issues.

\subsection{Naming a State}
\label{sec:state-name}

Over the course of the lifetime of a quantum network, many entangled
quantum states will be created and consumed.  Repeaters will make both
independent and coordinated decisions about which states to purify,
swap, error correct, forward, buffer, and discard, as they build
states that satisfy users' requests.  In order to communicate
successfully about these states, nodes must be able to \emph{name}
states using a namespace that other repeaters will understand: ``do
operation $U$ on this particular state we share.''  In order to
construct such requests unambiguously, the qubits within the states
must also be named.

The simplest naming scheme for a particular qubit is the tuple
$(N,A)$, where $N$ is the node name and $A$ is the physical qubit
address within the node.  However, there are three key problems with
this scheme:
\begin{itemize}
\item each node is entitled to move the logical state of a qubit from
  one physical qubit to another;
\item physical qubits are reused after being freed; and
\item the node issuing the original request may need to refer to the
  qubit by name even before physical resources for it are actually
  allocated (e.g., a request for a gate to be executed may be issued
  at the same time as the initial entangling pulse).
\end{itemize}
These factors mean that physical address is a constraining and
unreliable identifier for the quantum states that are our true subject
of interest.  All of these problems can be solved by allowing the
original requesting node to assign a \emph{virtual address} or other
abstract identifier for the qubit; the node (or network) housing the
physical resource is responsible for maintaining the mapping of
virtual to physical resource.  That mapping information is private to
the node and need not be disclosed or coordinated with other nodes.
In order to ensure that the virtual address assigned by the requester
is unique, the full address tuple must include the requesting node and
the actual request identifier.

The naming scheme must be prepared for names to shift as operations
proceed.  Multiple quantum states often merge to become a single
state.  Purification, entanglement swapping, and error correction all result in such
mergers flowing up the protocol stack, and result in multiple requests
moving down the protocol stack.  Names for states and qubits may be
remapped when crossing boundaries.  Names for nodes, when visible, do
not generally need to change, but requests moving from the outside of
a network to the inside may become more specific at the boundary.

Each boundary in the system, whether a software boundary between
modules or a hardware boundary between nodes or networks, represents a
point at which resource names and requests may change.  Logically,
these boundaries represent points where these mappings and requests
must be maintained, although in implementation this may vary.

\subsection{Defining Quantum Requests}

As classical distributed computation proceeds, applications running on
several nodes request that the network subsystem send and receive
messages or, using higher-level constructs, synchronize the state of
distributed copies of shared data
structures~\cite{coulouris:distributed-systems-4ed}.  In the quantum
world, a quantum request is for a specific state, spanning a named set
of nodes.

The interface to the network subsystem must allow the requester to
specify the desired state $|\psi_S\rangle$, while the network will
actually return
\begin{equation}
\rho = \trace_{AB} \ketbra{\Psi}{\Psi} \ \ \mathrm{where} \ \ \ket{\Psi} = \ket{\psi_S^\prime} \widetilde{\otimes} \ket{\psi_{A+B}}
\end{equation}
where $\ket{\psi_S^\prime}$ spans the set of state qubits, 
$\ket{\psi_{A+B}}$ is the set of ancillae (defined but unused
qubits, for this state) plus the bath (the environment), and
$\widetilde{\otimes}$ indicates that what we get in the real world is
only an approximation of a separable state.  The aim is to have
\begin{equation} \rho \approx \ketbra{\psi_S}{\psi_S} \end{equation}
\noindent within certain tolerances. The request must therefore also
specify these tolerances on the state: a minimum fidelity and a
maximum entanglement with the ancillae and bath.  Thus, a density
matrix should be viewed as \emph{no-less-than} for the element(s)
corresponding to the desired state, and \emph{no-more-than} for the
elements corresponding to undesired states.

Both the fidelity $F=\bra{\psi_S} \rho \ket{\psi_S}$ and entropy
$S=-\trace ( \rho \log \rho )$ appear in the request to constrain the
returned state $\rho$ to be near the desired state. The fidelity is to
ensure closeness to $\ket{\psi_S}$; the constraint on the entropy of
$\rho$ allows the system to filter out returned states that may be
non-trivially entangled with other nodes in the system.  In the limit
of $F\rightarrow 1$, the entropy becomes unnecessary, but for
fidelities bounded farther away from 1, the entropy becomes a useful
tool.  We assume that repeater nodes make repeated use of the same
physical resources, and sometimes swap data qubits with ancillae,
which if done imperfectly leaves behind some residual entanglement
between qubits which should not be entangled.  Further reuse of those
ancillae can therefore further entangle data qubits in an undesired
fashion.  Because both the qubits on which $\ket{\psi_S}$ are defined
and the ancilla qubits may be entangled with the environment (that is,
in a mixed state), the state of $\rho$ alone cannot determine if any
node qubits are entangled with any of the ancillae. Limiting the
entropy of $\rho$ serves to limit the possible entanglement with
ancilla qubits by limiting all external entanglement.

In addition to these properties, the requester must specify the
desired logical or physical encoding of the quantum state.  An
application will request an absolute encoding, while each layer in the
protocol stack provides a relative encoding (discussed further below),
with the entire stack to provide the absolute encoding.

The tuple specifying a request for a state is
\begin{equation}
T = (ID, |\psi_S\rangle, F, S, ((N_i,A_i)), E_A),
\end{equation}
\noindent where $ID$ is the transaction identifier assigned by the
requester, $F$ is the minimum acceptable fidelity of $\rho$ with
$\ket{\psi_S}$, and $S$ is the maximum acceptable entropy of
$\rho$. $((N_i,A_i))$ is the set of nodes that are requested to
comprise the state and the virtual addresses $A_i$ that are to be used
for the qubits, and $E_A$ specifies the absolute quantum error
correction encoding.  $|\psi_S\rangle$ is the desired pure state; the
exact encoding of the description of the requested state is beyond the
scope of this paper, but can take numerous forms, including state
vector, stabilizer, and circuit descriptions.

Requests may also be for \emph{actions} to be executed on specific
states, in which case the tuple is
\begin{equation}
T = (ID, C, F, S, ((N_i,A_i)), E_A),
\end{equation}
\noindent where $C$ is a circuit that may include both unitary and
measurement operations.

The return value of a request is the tuple
\begin{equation}
R = (ID, \rho)
\end{equation}
where $\rho$ is the density matrix of the delivered state for request
$ID$.  The set of resources represented by $\rho$ is specified by the
basis $((N_i,A_i))$, the tuple of tuples including node (or network)
identifiers $N_i$ and the virtual addresses $A_i$ included in the
original request.

Benjamin et al. described a brokered approach to building large-scale
graph states from smaller ones, tailored to a specific hardware
implementation~\cite{benjamin:brokered}.  QRNA provides a framework
for abstracting and generalizing this process, including support for
cost functions that will allow intelligent decisions for constructing
the sub-graphs.

\section{Implementing Recursion in Quantum Networks}
\label{sec:impl}

\subsection{Satisfying Quantum Requests}

Requests naturally originate at applications running on specific
nodes, and are processed through a series of software protocol modules
that implement the layers of the protocol stack, with carefully
defined interfaces between the layers.  Each layer in the protocol
stack has access to a set of resources it can use to satisfy requests:
it knows about a certain set of network nodes (or, more scalably, how
to \emph{find out} relevant information about a set of network nodes),
can ask for certain states (including entangled states) to be created
on that set of nodes and for certain operations to be performed on
those qubits, and can utilize its own internal capabilities.  It has
exclusive control of a certain set of resources, and may consult with
the corresponding layer instances at remote nodes about the best way
to satisfy requests.  However, it should endeavor to make
\emph{independent but coordinated} decisions whenever possible, so
that the latency penalty for explicit messaging can be avoided.

Each protocol instance has the ability to execute local quantum
operations (unitary operations and measurements), as well as compute
and communicate classically with other repeater nodes.  This ability
is often referred to as \emph{LOCC}, local operations and classical
communication.  The instance has no access to distributed quantum
states or operations beyond those it currently owns.  If additional
states are needed to complete an operation, they must be requested
from protocol layers below or from peers.

Requests are not constrained to be $1:1$; a single request from above
may be mapped to multiple requests to the layer below.  A protocol
layer has the right to merge and split states and issue multiple
requests to meet its obligations.  The ability to \emph{buffer}
quantum states, to hold them while waiting for other resources to
become available (e.g., other quantum states or answers to classical
queries), is generally necessary when coordinating multiple requests.

Protocols that make decisions about how to get from place to place in
the network must have access to a \emph{cost function} for specific
requests that can be used to make intelligent decisions, discussed
next.

\subsection{Finding Rendezvous Points}

Purify-and-swap repeaters require the explicit use of named rendezvous
points where the entanglement swapping occurs.  On a modest-sized
network, Dijkstra's algorithm~\cite{dijkstra1959ntp} can be applied to
select a path through the network, then the swapping points optimized
on the chosen chain of repeaters~\cite{satoh10:qdijkstra-aqis}.  The
order of entanglement swapping can be either specified or left
unspecified.

Store-and-forward repeaters need a similar path selection mechanism,
but have no direct need for waypoints, although they do require an
adequately scalable mechanism for calculating routes and choosing the
correct next hop on a request-by-request basis.  However, the Open
Shortest Path First (OSPF) protocol~\cite{rfc:ospfv2}, built on
Dijkstra's shortest path first algorithm, is widely used on the
Internet as an IGP in localized regions (called routing domains), but
its use is impractical for more than a few thousand nodes. The
Internet as a whole consists of tens of thousands of separate such
domains interconnected by a separate routing protocol (BGP, used as an
EGP), so that the hierarchy supports hundreds of thousands of
interdomain connections, with the entire Internet consisting of
perhaps tens of millions of routers and hundreds of millions of
end-nodes.  Hierarchy is the principal means of solving such
scalability problems, and the hierarchy and recursion allow the
details of the routing mechanism at each level to remain irrelevant as
long as the picture presented to the outside world is consistent.

\subsection{Example}
\label{sec:example}

\begin{figure}[t]
  \begin{center}
\includegraphics[height=50mm]{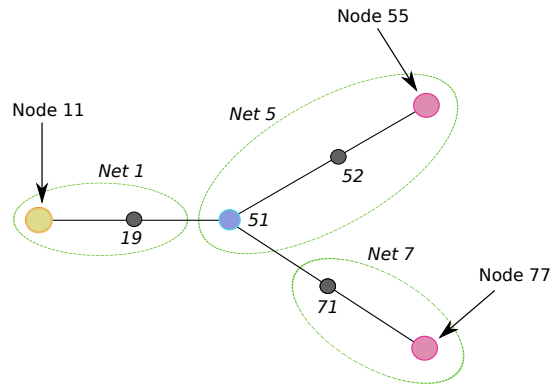}
  \end{center}
  \caption{Example of a small-scale internetwork composed of three
    networks.  Our example request is initiated at Node11, and
    includes Node55 and Node77.}
  \label{fig:small-network}
\end{figure}

\begin{figure}[t]
  \begin{center}
\includegraphics[height=50mm]{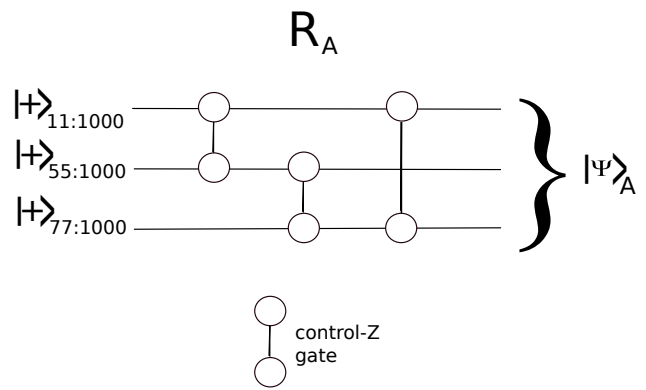}
  \end{center}
  \caption{Circuit for the three-qubit cluster state requested at
    Node11.  $11:1000$ and similar are the virtual addresses for the
    qubits, assigned by Node11.}
  \label{fig:app-circuit}
\end{figure}

As an example, consider an application needing a three-qubit cluster
state defined by the circuit in Fig.~\ref{fig:app-circuit}.  The
request originates at Node11, with the three qubits requested to be at
Node11, Node55, and Node77 in the network in
Fig.~\ref{fig:small-network}.  The application begins by specifying
the state it wants, then other (``system'') software running at Node11
creates a global strategy for how to achieve the state, and sends
requests to corresponding nodes or networks.  The bulk of this work
happens in the QRNA {\tt process()} step in Fig.~\ref{fig:algorithm}.
The nodes that receive the requests will in turn will craft their own
strategies for the requests they receive.  Although the two stages of
creating a strategy and choosing where to send the application
requests are intertwined, here we will describe them separately for
clarity.

The application running on Node11 creates a request of the form\footnote{These example request values are taken from the output state of purification in a quantum network simulator \cite{ladd06:_hybrid_cqed}.}
\begin{eqnarray}
R_A & = & (1, \ket{\psi_A}, F\ge 0.99, S\le 0.1, \nonumber\\
{} & {} &((\textrm{Node11},1000),(\textrm{Node55},1000), \\
{} & {} & (\textrm{Node77},1000)), \textrm{Raw}), \nonumber
\end{eqnarray}
where $\ket{\psi_A}$ is the cluster state created by the circuit in
Fig.~\ref{fig:app-circuit}, $1000$ is the virtual address chosen to be
used for the qubit requested at each node, and Raw indicates that we
are requesting an unencoded state.

To fulfill $R_A$, the first system software module to process the
request (still at Node11) must create a global strategy.  The
principal decision is whether to create the state in one location and
move the qubits via
teleportation~\cite{bennett:teleportation,furusawa98,S.Olmschenk01232009},
or allocate the qubits in place and use teleported gates to execute
the circuit in a remote fashion~\cite{gottsman99:universal_teleport}.
For our circuit, either approach results in three remote operations.
The exact cost of a teleportation or a remote gate will depend on the
network; we defer discussion of the detailed cost model and decision
function to future work.  We will consider the case where Node11
decides to ask for the state to be created in one location, then
propagate the component qubits outward to the requested nodes.

With the global strategy chosen, the next step is selecting
\emph{where} each of the operations will take place.  The routing
table at Node11, shown in Table~\ref{tab:example-rtg-tab-11}, contains
information on how to get to all destinations on the network.  To
achieve scalability, the table has more precise information about
nearby destinations, and vague information about more remote
destinations, achieved using hierarchy and recursion, as in classical
networks~\cite{tanenbaum:networks-4th-ed}.

The most important question is where to build the cluster state.
Based on a cost function that uses the information in the routing
table, Net5 is identified as being close to the ``center'' of this
request.  Thus, the strategy module chooses to ask Net5 to create the
state, after which Net5 will teleport the qubits to Node11, Node55,
and Node77.

\begin{figure*}[t]
  \begin{center}
    \includegraphics*[width=100mm]{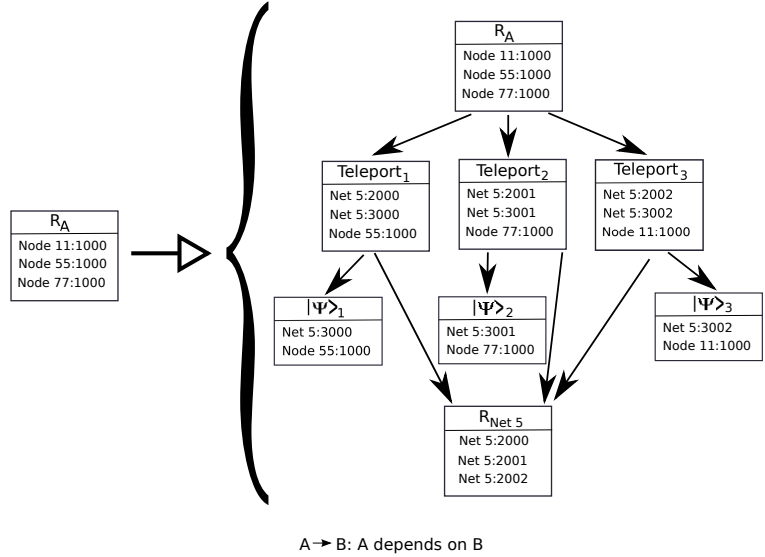}
  \end{center}
  \caption{The initial application request $R_A$ is translated into
    a set of requests for sub-operations before leaving its origin,
    Node11.  Each box lists not the full request tuple, but only the
    ID of a sub-request and the virtual addresses for the qubits
    assigned by the request creator.}
  \label{fig:broken-down}
\end{figure*}

As shown in Fig.~\ref{fig:broken-down}, the original request (left
side of the figure) is broken down into seven separate requests (right
side of the figure): one for the state to be created local to Net5
(labeled $R_{Net5}$), three for Bell pairs to be used for
teleportation (labeled $\ket{\Psi}_1$, etc.), and the teleportation
operations themselves.  In this case, $R_{Net5}$ is the
same circuit as in Fig.~\protect\ref{fig:app-circuit}, with the
resources specified as local to Net5 rather than distributed.

Each box in the figure lists the virtual addresses of the qubit
resources to used for that request.  The virtual addresses are created
when the requests are created, but are not assigned to matching
physical resources until the requests are processed at the receiving
nodes.  Each of these requests must also carry information about
fidelity and entropy, with those values chosen to ensure that the
delivered final state will meet the originally-requested constraints.
Based on the routing table in
Tab.~\protect\ref{tab:example-rtg-tab-11}, each of these requests is
then sent via the classical network to each node involved; in this
case, Node51, as the gateway to Net5, will receive most of the
requests.  Node51 will then forward the requests onward, or craft its
own strategy, as appropriate.  Requests can be executed once all
dependencies (indicated with arrows in Fig.~\ref{fig:broken-down}) are
satisfied.  The application's request is completed once all of the
component requests finish.

Although this example shows only a single layer of recursion, the
process may be repeated indefinitely for the physical nodes (as shown
in Fig.~\ref{fig:network}), or for requests.  To achieve adequately
high fidelities, the node assigned to process each of these requests
may in turn break the request down further into multiple requests for
base-level entangled Bell pairs and purification operations.
Likewise, for those operations spanning multiple hops, either
entanglement swapping or hop-by-hop teleportation can be requested.

\begin{table}[t]
  \caption{The routing table at Node11 contains information on how to
    get to all destinations on the network.  To achieve scalability,
    the table has more precise information about nearby destinations, and
    vague information about more remote destinations, achieved
    using hierarchy and recursion.  Node55 resolves to Net5, and
    Node77 resolves to Net7, so that independent records are not
    needed for each node.}
\label{tab:example-rtg-tab-11}
\begin{center}\small
\def\arraystretch{1.2}
\begin{tabular}{|c|c|}\hline 
Destination & Route \\ \hline \hline
Node19 & (direct) \\ \hline
Net1  & Local       \\ \hline
Net5  & Node19       \\ \hline
Net7  & Net5       \\ \hline
\end{tabular}
\end{center}
\vspace*{-4mm}
\end{table}

\begin{table}[t]
  \caption{The routing table at Node51.}
\label{tab:example-rtg-tab-51}
\begin{center}\small
\def\arraystretch{1.2}
\begin{tabular}{|c|c|}\hline 
Destination & Route \\ \hline \hline
Node52 & (direct)  \\ \hline
Node55 & Node52  \\ \hline
Net1 & Node19 \\ \hline
Net5  & (process locally) \\ \hline
Net7  & Node71 \\ \hline
\end{tabular}
\end{center}
\vspace*{-4mm}
\end{table}

\section{Conclusions}

The fundamental difference between classical and quantum networks is
the services they deliver.  Classical networks move data from a source
application to one or more destination applications over a distance.
Quantum networks may likewise transport data from place to place, but
in addition can produce distributed entangled quantum states,
connecting two or more quantum applications.  This difference requires
a new form of interaction between network components.  On the
Internet, a received packet is implicitly a request: please forward
this data toward the destination or destinations listed.  In our
Quantum Recursive Network Architecture (QRNA), rather than such an
implicit request, the requester explicitly asks a node or network to
participate actively in the creation of a larger state.  Thus, rather
than simply an information transfer system, a quantum network is a
general-purpose distributed quantum computing system.

The problems of truly large-scale quantum repeater networks have much
in common with the problems of classical distributed computing: naming
and resource management are critical issues, and judicious use of the
concepts of hierarchy and recursion provide the right abstraction to
keep the systems efficient while the data structures that must be
managed at each node remain tractable in size.  Dynamic composition of
the protocol stacks provides the needed flexibility, as well as
isolation of responsibility.

All of these issues can be addressed through the use of recursive
networking.  QRNA abstracts subnetworks as individual nodes, allowing
technology-independent requests for quantum state creation to be
constructed with imperfect knowledge of the total network structure
and state, and for those requests to be modified and processed in a
recursive fashion as necessary to deliver the end-to-end quantum state
required by applications running on quantum computers.

This paper has assumed that network nodes and repeaters are
well-behaved and are not malicious, but in the real world those
assumptions will not hold and the issues of robustness in the
sometimes-hostile world will have to be addressed.

Long computations will naturally require not a single distributed
state, but a sequence of them; reservations for such longer sequences,
especially the real-time requirements, are beyond the scope of the
current discussion.

Although we have focused in this paper on the creation of a core group
of entangled states that are common building blocks for distributed
algorithms, the mechanisms generalize quite easily to support direct
distributed execution of any quantum algorithm.

We expect that adoption of QRNA will provide operational benefits and
reduce redundant engineering effort as quantum repeater networks
evolve and grow.

\noindent{\sf Acknowledgement}\quad RV and CH thank Bill Munro for useful discussions.
RV acknowledges funding from JSPS
KAKENHI.  CH is supported by the Japan Society for the Promotion of
Science (JSPS) through its ``Funding Program for World-Leading
Innovative R\&D on Science and Technology (FIRST
Program)''. This work was partly supported by the NSF (Grant
No. CNS-0626788). Any opinions, findings, and conclusions or
recommendations expressed in this material are those of the authors
and do not necessarily reflect the views of the National Science
Foundation.\\

\bibliographystyle{unsrt}

\end{document}